# Eu valence transition behavior in the nano form of EuPd$_2$Si$_2$

Kartik K Iyer, Tathamay Basu, P.L. Paulose, and E.V. Sampathkumaran
*Tata Institute of Fundamental Research, Homi Bhabha Road, Colaba, Mumbai-400005, India*

**Abstract**

The compound EuPd$_2$Si$_2$ is a well-known fluctuating-valent compound with a largest variation of Eu valence in a narrow temperature interval (around 150 K). The ball-milled form of this compound was investigated to understand the Eu valence behavior in the nanoform. The compound is found to retain the ThCr$_2$Si$_2$-type tetragonal structure after ball-milling leading to a reduction in particle size, typically falling in the range 10 – 100 nm. We find that there is a qualitative change in the temperature dependence of magnetic susceptibility for such small particles, with respect to that known for bulk form. To understand this microscopically, Mössbauer spectra as a function of temperature were taken. The Mössbauer spectrum of the nanocrystalline compound is essentially divalent-like at room temperature, but becomes distinctly bimodal at all temperatures below 300 K, unlike that of the bulk form. That is, there is a progressive transfer of intensity from divalent position to trivalent position with a gradual decrease of temperature. We attribute it to a first-order valence transition, with extreme broadening by defects in the nano specimen. Thus there is a qualitative change in the valence behavior in this compound as the particle size is reduced by ball-milling. Such a particle size study is reported for the first time for a Eu-based mixed valent compound.

Key words: EuPd$_2$Si$_2$, Intermediate valence, ball-milling, nano form, first-order

PACS numbers: 71.28.+d; 71.20.Eh; 73.61.At; 73.20.Jc

**Introduction**

The interest in the study of the physics of rare-earths became exciting nearly half a century ago following the discovery of the fact that the 4f occupation number (in other words, valence) for some rare-earths (Ce, Sm, Eu, Yb) can fluctuate under favorable circumstances due to quantum mechanical mixing of 4f orbital and the conduction band [1]. The valence fluctuation (also called intermediate valent, IV) behavior of the Eu-based intermetallics are often found to be different from that encountered among Ce and Yb-based materials in the sense that the net change in the valence in the former is in general significantly large compared to that in the latter. That is, the valence for Ce generally does not exceed 3.3 in a metallic environment [2], whereas, for Eu compounds, for instance, in $EuCu_2Si_2$ – the first Eu-based exotic IV compound reported as early as 1973 [3] - the mean valence was shown to vary by as much as 0.8 [4] with temperature ($T$). This report [3] prompted numerous groups to probe the valence transition behavior of Eu extensively in many metallic systems, bringing out such dramatic changes, not only as a function of $T$, but also as a function of magnetic field [$H$] [6], pressure, and chemical doping [see, for instance, Refs. 5 – 10]. However, to our knowledge, there is no report on the behavior of Eu valence in such exotic compounds when the particle size is reduced to nano form, despite the fact that the nanoscience area of research has been at the forefront of modern science.

We have therefore attempted to probe the Eu valence behavior of $EuPd_2Si_2$ [11, 12] in the nanoform, synthesized by high-energy ball-milling, as a continuation of our similar efforts on Ce based intermetallics [13]. This compound, forming in the $ThCr_2Si_2$-type tetragonal structure (space group I4/mmm), was originally pioneered [12] to exhibit the largest and the most rapid variation of mean valence of Eu (as measured by $^{151}$Eu Mössbauer isomer shift) in a narrow temperature interval (130-170 K). In this respect, this compound remains to be the most exceptional stoichiometric IV compound of Eu, though a small chemical doping (Pt, Au, or Sn for Pd) was subsequently claimed to induce first-order variations of Eu valence [14], the origin of which was not clear. It may be added that such sharp valence changes were later reported in some other isostructural Eu-based pseudoternary materials [5, 15] as well. Due to this reason, this compound was studied by several experimental methods for the past three decades, including a very thorough work comparing the valence obtained by Mössbauer and x-ray absorption techniques [16]. It is also intriguing to note that the application of a magnetic field of about 90 kOe at 6 K has been shown to transform the nearly trivalent state to the nearly divalent state through a first-order transition, due to the Zeeman energy gain being larger than the energy difference between the two valence states of Eu [17]. Clearly, the interest in this ternary family of Eu-based materials among solid state physicists remain unabated for the past five decades. In this article, we report the results of dc magnetization ($M$) and $^{151}$Eu Mössbauer effect measurements on the ball-milled specimens of $EuPd_2Si_2$ for the first time.

**Experimental details**

The polycrystalline form of the sample was initially made by arc melting stoichiometric amounts of constituent elements (purity >99.9% for Eu, and >99.99% for Pd and Si) in an arc furnace in argon atmosphere. In order to compensate for the loss of Eu while melting, an additional amount of Eu (10 %) was added. The ingot was then annealed at 900 C for 340 h. The sample thus obtained was found to be single phase by x-ray diffraction (XRD) (Cu K$_\alpha$). The bulk form thus obtained was ground for 2.5 h in an atmosphere of toluene in a high-energy ball-mill and the specimens were removed intermittently (after 0.5, 1, 1.5 and 2.5 h) for XRD studies. A transmission electron microscope (TEM) (Tecnai 200 kV) was used to understand the reduction in particle size following milling for 2.5 h. The dc magnetization studies (2 – 300 K) were carried out by a commercial SQUID magnetometer (Quantum Design). $^{151}$Eu Mössbauer spectra were obtained at selected temperatures (2 to 300 K) using the $^{151}$SmF$_3$ source (E$_\gamma$= 21.6 keV). Magnetization and Mössbauer studies were performed for the specimen milled for 2.5 h only and not for those with intermittent grindings.

**Results and discussions**

The XRD patterns are shown in figure 1 and the linewidths are found to increase with milling time, which is consistent with the reduction in the particle size due to milling. The TEM picture for the specimen milled for 2.5 h, shown in figure 2, reveals the presence of very small particles, e.g., 10nm, with a spread in particle size. Therefore, we determined the average particle size from the width of the XRD lines employing Debye-Scherrer formula and this analysis yields a value of about 100 nm for the particle size. In short, there is a large spread in the size of the particles (typically 10 – 100nm). The lattice constants are found to undergo a marginal increase with the reduction in particle size and we attribute it to the tendency of the valence of Eu to move closer towards 2+.

Magnetic susceptibility ($\chi$) measured in a magnetic field of 5 kOe is shown in figure 3 for the milled specimen. We have also measured $\chi$ for the bulk specimen (which was used for milling) and it was found that there is a peak followed by a drop around 150 K as reported in the past literature [12] with decreasing temperature due to the change of valence of Eu from the high-moment state (corresponding to a mean valence close to 2) to a low-moment state (corresponding to a mean valence close to 3). There is an upturn at much lower temperatures, which has been known to arise due to a small amount of Eu remaining in 2+ state as a result of certain degree of crystallographic (Pd-Si) disorder and the fraction of this component has been known to vary one sample to another [16]. It is transparent from this figure that the $\chi$ values are very close for the bulk and nano form near room temperature, but the curve for the latter tends to lie higher in the plot with a gradual lowering of temperature. This implies that the average valence of Eu in both the forms is nearly the same near room temperature. Notably, the peak around maximum is absent in the nano form and $\chi$ is found to undergo continuous increase with decreasing temperature down to 2 K. Thus, there is a qualitative change in the shape of $\chi(T)$ curve due to milling. Inverse $\chi$ exhibits linear variation above ~150 K (see figure 3, bottom). The effective moment obtained by Curie-Weiss fitting of this linear region turns out to be ~7.6 $\mu_B$/Eu, which is less than the theoretical value (7.94 $\mu_B$/Eu) for divalent Eu; this indicates that Eu is intermediate valent even in the nano form (with a small admixture from trivalent state) in this high-T range. There is a deviation from the high temperature linearity, the explanation for which can be obtained after observing Mössbauer spectral features (see below).

$^{151}$Eu Mössbauer spectra of the ball-milled specimen, recorded at various temperatures are shown in figure 4. At 290 K, the spectrum appears to be a superposition of two components. There is a strong resonance with an isomer shift close to -8 mm/s, which could be fitted to a Lorenzian form. The magnitude of the isomer shift is marginally higher than that observed (-7.4mm/s) for the bulk form [12]. [For the benefit of the readers, we recall that the isomer shift values for divalent and trivalent Eu are known to fall typically in the range -8 to -13 mm/s and 0 to 3 mm/s respectively]. This value establishes that the valence of Eu for the majority of Eu ions is relatively more towards 2 for the nanoform at room temperature. In addition, there is a weak component around 1 mm/s, as known for the bulk form, suggesting the existence of a small fraction of nearly trivalent component. The linewidths at half maximum (LWHM) are also significantly enhanced (about 4mm/s) which is typical of small particle size. In figure 4, the continuous lines through the data points are obtained by a simplified fitting to two Lorentzian lines at these two positions, though it is difficult to rule out the existence of resonances with slightly different isomer shifts for each of these components, depending on the local environment or particle sizes. It is transparent from this figure that, as the temperature is gradually lowered, the relative intensity (obtained from the areas of the subspectra) of the nearly trivalent component increases at the expense of the component near divalent position and it is quite significant below about 150 K. At 40 K, these two components are of comparable intensity. The intensity ratio and the isomer shifts of the two lines are shown in figure 5. It is to be noted that the isomer shifts of both the features, obtained by above mentioned fitting, change with lowering temperature and the magnitude of this shift is larger than the well-known thermal red shift (<0.1mm/s), as though both the components are in an intermediate valent state. Interestingly, the magnitude of the isomer shift of the divalent component becomes larger in magnitude, towards more divalent character, with decreasing temperature, as reported for the divalent satellite component for the bulk form in the literature [16]. The situation is somewhat similar for the feature around 0 mm/s, however with an unusual tendency to deviate from trivalency towards divalency at lower temperatures. It is also intriguing that the T-dependence of isomer shift for the nearly trivalent component is somewhat non-montonous with a minimum and the exact origin for this is not clear; possibly, the changes in the electronic structure and/or phonon density of states [10] in different regions of the specimen following above-mentioned intensity-transfer could be responsible for this. It is of great interest to address this aspect

theoretically. Both the features show large changes around 150 K, which could be responsible for the deviation from the high temperature Curie-Weiss behavior of χ around this temperature. It should however be stressed that there is no gradual movement of the position the divalent component towards trivalent position, unlike in bulk form [as shown in in the inset of figure 5a; References 12, 16]; that is, there is no resonance with intermediate values of isomer shift, say, -5mm/s, as observed for the bulk form at 200 K. Thus, there is a gradual dumping of intensity from nearly divalent position to nearly trivalent position with decreasing temperature. On the basis of this, we suggest that there is a sharp change in the valence of Eu from a nearly divalent state to nearly trivalent state for a certain fraction of Eu ions at a given temperature and that there is a very large spread in transition temperatures depending on the local chemical environment, that is, due to defects/disorder around a given Eu ion in the nano form. This kind of 'dumping of intensity' rather than gradual movement of the resonance position was in fact observed by a small chemical doping in $EuPd_2Si_2$ [14]. The fact that we are able to observe this in an undoped specimen implies that the defects due to doping, rather than the presence of the dopant, could induce this 'intensity dumping' feature. Finally, we could fit the 'untransformed' divalent component below 8 K to a magnetic hyperfine splitting, suggesting magnetically ordered state for these Eu ions at low temperatures. This aspect for such a 'satellite feature' was well-studied for the bulk form by Wortmann et al [16] and the magnetic hyperfine split Mössbauer spectra revealed the existence of magnetic ordering of these ions well below 10 K. It is interesting to note that Mδssabuer spectral features, including the isomer shift behavior, of such a satellite are not modified by milling. A careful look at the inverse χ curve in figure 3 (bottom) reveals a change in the curvature below about 8 K, tending to a flattening, as a signature of possible magnetic ordering of this component.

**Conclusion**

We have studied the compound $EuPd_2Si_2$ – which remains to be an exotic intermediate valent system of Eu even after its discovery several decades ago – in its nano form obtained by ball-milling. Such a study in nanoform for any Eu intermetallics is reported for the first time. We find considerable differences in the Eu valence behavior with respect to the bulk form. That is, a sharp change of intermediate valence value of Eu ion in the range 100-150 K from nearly divalent state near 300 K to nearly trivalent state near 50 K known for the bulk form is not apparent in the magnetic susceptibility data of the ball-milled specimens. Instead, on the basis of Mössbauer spectral intensity behavior, we conclude that this transition is abrupt for certain fraction of Eu ions at any given temperature with a large spread in the transition temperature. The local environmental effects and defects in the nano particles apparently play a decisive role on this behavior. We therefore believe that the first-order changes in the valence induced by marginal chemical doping in the bulk form of $EuPd_2Si_2$ also could be due to defects induced by such doping.

The authors thank B.A. Chalke and R.Bapat for TEM analysis.

Figure captions:

**Figure 1:** X-ray diffraction patterns of ball-milled $EuPd_2Si_2$ specimens.

**Figure 2:** Transmission electron spectroscopic images of the 2.5h milled specimen of $EuPd_2Si_2$.

**Figure 3:** (Top) Magnetic susceptibility as a function of temperature, measured in a field of 5 kOe, for the bulk and 2.5h milled specimens of $EuPd_2Si_2$. (Bottom) Inverse susceptibility for the milled specimen and the continuous line through the high temperature points is a fit to the Curie-Weiss form.

**Figure 4:** $^{151}$Eu Mössbauer spectra of 2.5 h ball-milled specimen of $EuPd_2Si_2$. The lines through the data points is a fit to two Lorentzian forms, one near divalent position and the other near trivalent position.

**Figure 5:** Temperature dependence of (a) isomer shift of nearly divalent and nearly trivalent Eu components, and (b) intensity ratios of the two components, for 2.5 h ball-milled $EuPd_2Si_2$. The lines through the points serve as guides to the eyes. The temperature dependence profile of isomer shift of main Eu Mössbauer resonance, as reported in the literature, is reproduced in the inset of figure (a).

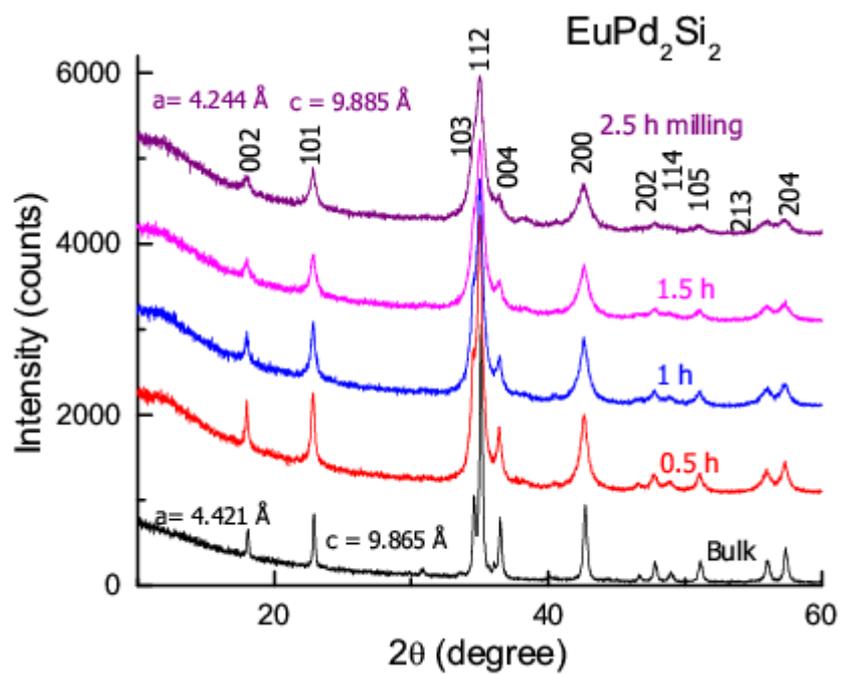

Figure 1
Figure 2

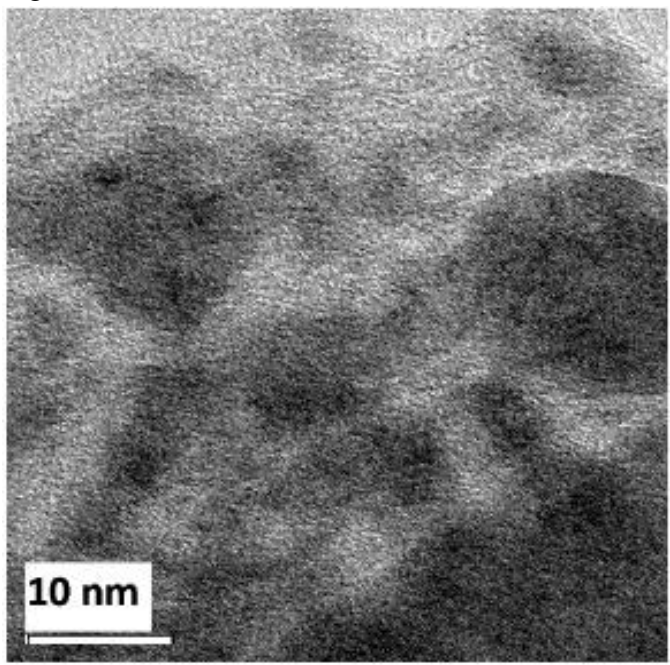

F10 nm

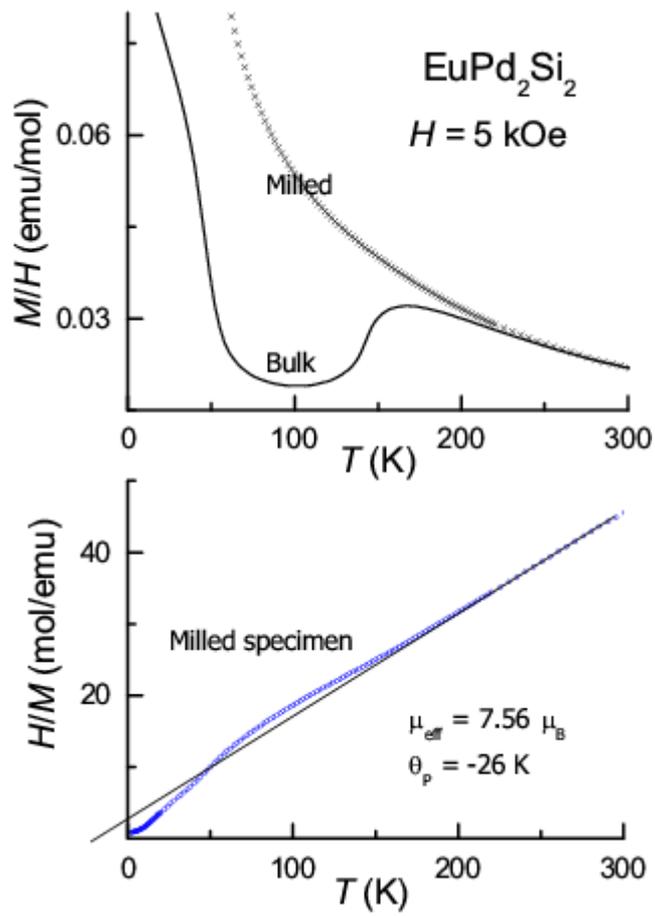

Figure 3

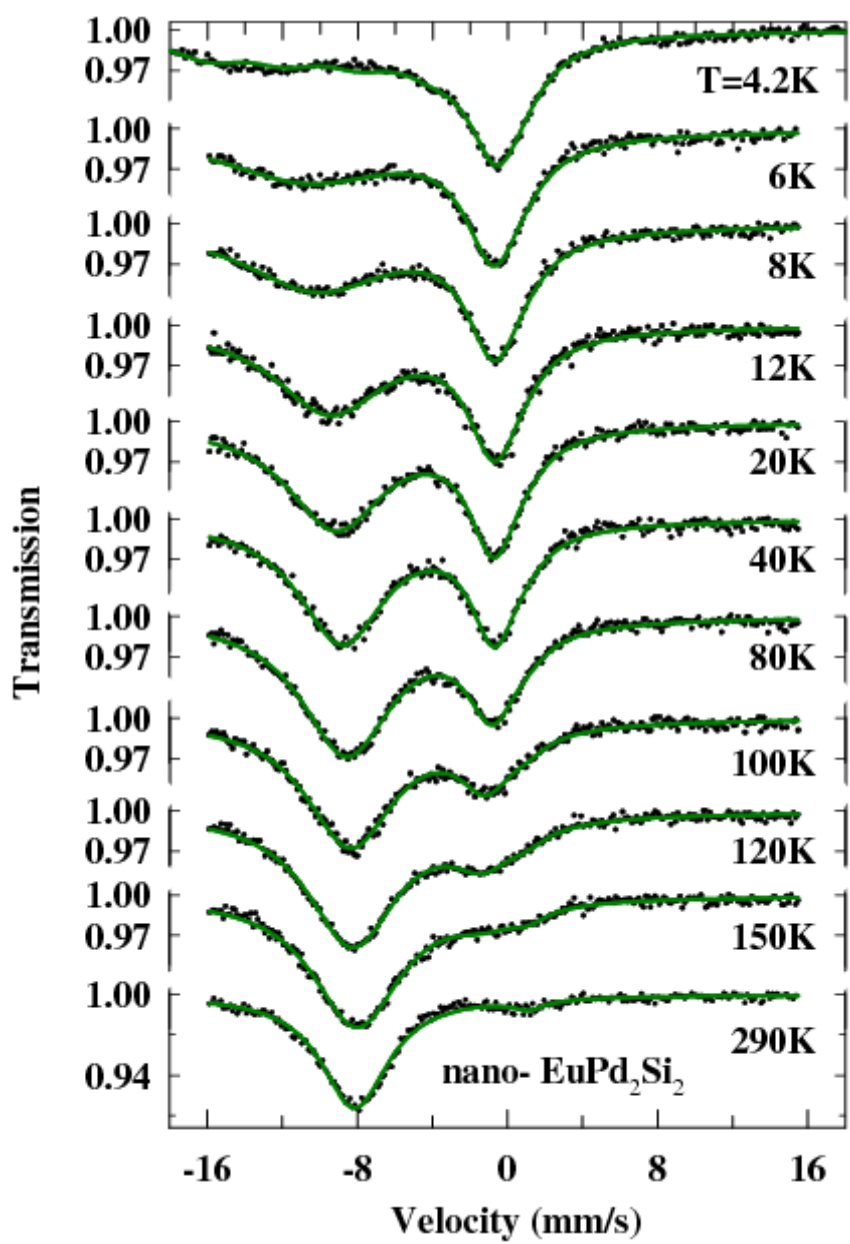

Figure 4

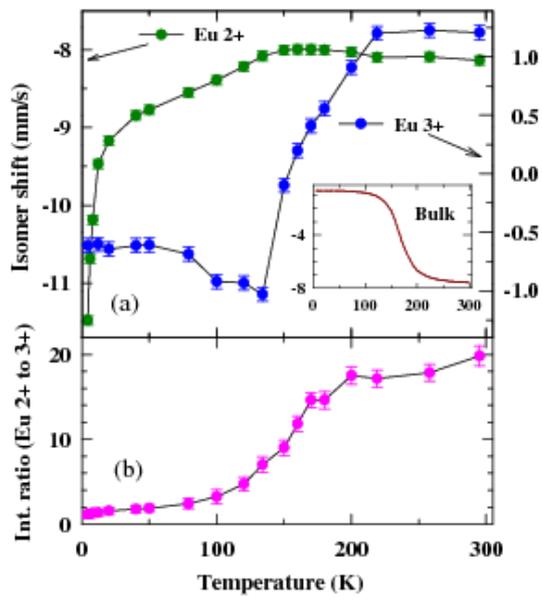

Figure 5.